\documentstyle[aps]{revtex}
\begin{document}
\voffset=-2mm
\title{Evolution of One--Particle and Double--Occupied Green Functions 
for the Hubbard Model at Half--Filling With 
Lifetime Effects Within The Moment  
Approach.}
\author{\it S. Schafroth}
\address{Physik-Institut der Universit\"at Z\"urich, 
Winterthurerstrasse 190, CH-8057 Z\"urich, 
Switzerland\\e-m: schafrot@physik.unizh.ch}
\author{\it J.J. Rodr\'{\i}guez--N\'u\~nez}
\address{Departamento de F\'{\i}sica-CCNE, 
Universidade Federal de Santa Maria,   
97105-900 Santa Maria/RS, 
Brazil. \\e-m: jjrn@ccne.ufsm.br}
\date{\today}
\maketitle

%
%
\begin{abstract}
We evaluate the one--particle and double--occupied Green functions for the 
Hubbard model at half--filling using the moment approach of 
Nolting\cite{1}. Our starting point is a self--energy, 
$\Sigma(\vec{k},\omega)$, which has 
a single pole, $\Omega(\vec{k})$, with 
{\it spectral} weight, $\alpha(\vec{k})$, and 
quasi-particle lifetime, $\gamma(\vec{k})$\cite{9709080}. 
In our approach, $\Sigma(\vec{k},\omega)$ becomes the 
central feature of the many--body problem and due to 
three unkown $\vec{k}$--parameters we have to satisfy 
only the first three sum rules instead of four as in the 
canonical formulation of Nolting\cite{1}. 
This self--energy  choice forces our system to 
be a non--Fermi liquid for any 
value of the interaction, since it does not vanish at 
zero frequency. The one--particle Green function, 
$G(\vec{k},\omega)$, shows 
the finger--print of a strongly correlated system, i.e., a 
double peak structure in the one--particle 
spectral density, $A(\vec{k},\omega)$,  
vs $\omega$ for intermediate values of the interaction. Close 
to the Mott Insulator--Transition, $A(\vec{k},\omega)$, becomes 
a wide single peak, signaling the absence of quasi--particles. 
Similar behavior is observed for the real and imaginary parts 
of the self--energy, $\Sigma(\vec{k},\omega)$. The 
double--occupied Green function, $G_2(\vec{q},\omega)$ 
has been obtained from $G(\vec{k},\omega)$ by means of the 
equation of motion. The relation between $G_2(\vec{q},\omega)$ 
and the self--energy, $\Sigma(\vec{k},\omega)$, is  
formally established and numerical results for the spectral 
function of $G_2(\vec{k},\omega)$, $\chi^{(2)}(\vec{k},\omega) 
\equiv -\frac{1}{\pi}\lim_{\delta \rightarrow 0^+}
Im\left[G_2(\vec{k},\omega)\right]$, are 
given. Our approach represents the simplest way to include: 
{\bf 1-} lifetime effects in the moment approach of 
Nolting, as shown in the paper; {\bf 2-} Fermi or/and 
Marginal Fermi liquid features as we discuss in the conclusions.\\
\\
Pacs numbers: 74.20.-Fg, 74.10.-z, 74.60.-w, 74.72.-h
\end{abstract}


%
%
\section{Introduction}\label{sec:intro}

	After the discovery of the High-$T_c$ materials\cite{3}, 
the study of correlations has gained interested due 
to the fact that there is the belief\cite{4} 
that the normal properties of these materials could 
be explained in the framework of the Hubbard 
model\cite{5,6}, since electron correlations 
are strong, i.e., the on-site electron-electron 
repulsions $U$ are much larger than the energies 
associated to the hybridization of atomic orbitals 
belonging to different atoms\cite{7}. We consider the study 
of correlations in the Hubbard 
model as a rewarding task since it will shed light on still unsolved points of
the novel materials. For example, at high temperatures ($T_c~30-130~K$) these
HTSC cuprates, which are poor conductors, become superconductors. 
This feature is strange indeed because the Coulomb repulsion is
strong. Contrary to the 
predictions of the Fermi liquid theory, the resistivity at $T~>~T_c$ 
and optimum doping is linear in temperature, i.e., 
$R~\approx~T$\cite{ShenDessau}.  
This suggests a very strong
scattering of elementary excitations. A discussion of the possible 
breakdown of Fermi liquid theory is given in Ref.\cite{Alexei}. In the 
present work we explore the effects of having a non--Fermi liquid 
behavior into the one--particle and double-occupied  
dynamical properties.

\indent We will use the moment approach 
(or sum rules) of Nolting\cite{1} for the spectral density,  
$A(\vec{k},\omega)$. It is well known in the literature\cite{ours} that 
the moment approach in the spherical approximation - when 
the narrowing band factor, $B(\vec{k})$, is not $k$ - dependent - 
{\it always}  gives a gap in the density of states $(DOS)$. 
If the chemical potential happens to be in this gap, then we {\it 
always} have an insulator. It has been argued that the way to cure 
this unrealistic gap is to have a better approximation for the 
narrowing band factor, $B(\vec{k})$. 

\indent We have followed a different path 
which consists in proposing a single pole structure in the self-energy,  
$\Sigma(\vec{k},\omega)$. Closing the gap for small and intermediate 
values of $U/W$ ($W$ is the bandwidth and $W = 8t$ in two dimensions), 
is not the only rationality behind our calculation. It is well 
documented\cite{Fukuyama,Andplus,Narikiyo} that 
correlations give rise to lifetime 
contributions that conspire against the very definition of quasi--particles 
at the chemical potential. This leads to an eventual collapse of the 
Fermi liquid picture of adiabaticity between the electron gas and 
the electron liquid so well described by the phenomenological 
treatment of Landau.
	
	This paper is organized as follows. In Section \ref{II}, we 
present our model Hamiltonian and our self--energy proposal, 
in which we try to justify our bold choice, as one step forward to understand 
the physics behind the moment approach of Nolting. This led us to 
come up with an Ansatz which is not zero at zero frequency, i.e., 
our system is not a Fermi liquid for any value of the interaction. 
We could argue that our Ansatz is valid for energy scales not too 
close to the chemical potential. However, our Ansatz has to become 
exact in the limit of $U/W \gg 1$. 
In Section \ref{III}, we present our results which consist of the 
real and imaginary parts of the  self--energy, and spectral functions 
of the one- and two--particle Green functions along the diagonal of 
the Brillouin zone, for different values of the interaction. 
We have used a system size of $32\times 32$. In 
Section \ref{IV} we present our conclusions and the future trends.

\section{Model Hamiltonian and Self-Energy Ansatz}\label{II}

\indent The model we study is the  Hubbard Hamiltonian
\begin{eqnarray}\label{Ham}
H = t_{\vec{i},\vec{j}}
	c_{\vec{i}\sigma}^{\dagger}c_{\vec{j}\sigma}
   + \frac{U}{2} n_{\vec{i}\sigma}n_{\vec{i}\bar{\sigma}}   
   - \mu c^{\dagger}_{\vec{i}\sigma}c_{\vec{i}\sigma}~~,
\end{eqnarray}
where $c_{\vec{i}\sigma}^{\dagger}$ ($c_{\vec{i}\sigma}$) are creation
(annihilation) electron operators with spin $\sigma$. $n_{\vec{i}
\sigma} \equiv c_{\vec{i}\sigma}^{\dagger}c_{\vec{i}\sigma}$. 
$U$ is the local interaction, $\mu$ the chemical 
potential and we work in the grand canonical ensemble. We have 
adopted Einstein convention for repeated indices, i.e., for the $N_s$ 
sites $\vec{i}$, the $z$ nearest-neighbor sites $(n.n.)$  
$\vec{j}$ and for spin up and down ($\sigma = -\bar{\sigma} 
= \pm 1$). $t_{\vec{i},\vec{j}} = -t$, for $n.n.$ and zero otherwise.

\indent Let us propose for the self--energy, 
$\Sigma(\vec{k},\omega)$, the following single pole Ansatz
\begin{equation}\label{single}
\Sigma(\vec{k},\omega) = \rho U + \frac{\alpha(\vec{k})}{\omega - 
\Omega(\vec{k}) - i\gamma(\vec{k})}~~~ ; ~~~ 
\alpha(\vec{k}),\gamma(\vec{k}) \in \Re ~~~   
\end{equation}
With our choice for $\Sigma(\vec{k},\omega)$, we have that the real 
and imaginary parts of the self--energy {\bf do} satisfy the 
Kramers--Kronig relations\cite{CL}, since it is analytic in 
one of half of the complex plane. In fact, the physical solution 
to the problem is when $\alpha(\vec{k}) \geq 0$, as it can be 
checked by finding the roots of $G(\vec{k},\omega)$ in the 
complex plane. We postpone the discussion of 
this point for the conclusions. The Ansatz given in Eq. (\ref{single}) 
has some similarity with the Hubbard--I solution\cite{HubbardIII}. 
However, we have neglected any frequency dependence in the 
damping. Our calculations show that $\gamma(\vec{k})$ is 
$\vec{k}$--independent but strongly $U$--dependent.

\indent The 
validity of Luttinger theorem\cite{Luttinger} has been 
discussed in Ref.\cite{shock}. We argue that most 
likely the Luttinger theorem is not going to hold because we have 
a non--Fermi liquid system. Our choice of $\rho U$ in Eq. (\ref{single}), 
the Hartree--shift, is very convenient since it redefines 
an effective chemical 
potential, $\mu_{eff} = 
\mu - \rho U$. This effective potential is zero at half--filling, 
$\rho = 1/2$, since $\mu = U/2$ there. Then, $\omega = 0$ 
means that we are at the 
chemical potential. We want to explicitly state 
that our choice (Eq.  (\ref{single})) 
has the advantage of requiring only three sum rules to 
be satisfied, instead of 
four as in the now normal procedure of Nolting\cite{1} which starts from the 
spectral function, $A(\vec{k},\omega)$, itself.  
We add that if $\gamma(\vec{k}) = 0$,  
then we go back to Nolting's canonical results. In this case\cite{RNMM}
\begin{equation}\label{spherical}
\alpha(\vec{k}) = \rho (1-\rho)U^2 ~~~ ; ~~~ \Omega(\vec{k}) = 
(1-\rho)U + B(\vec{k}) ~~~ ,~~~ B(\vec{k}) = B + F(\vec{k})~~~,
\end{equation}
\noindent where $B(\vec{k})$ is 
the narrowing band factor. The $\vec{k}$--independent 
narrowing band factor, $B$, is calculated in 
closed form in Ref.\cite{1}. It is a self--consistent 
quantity, though. The 
$\vec{k}$--dependent narrowing band factor, $F(\vec{k})$, has been 
evaluated recently by Herrmann and Nolting\cite{HN} using a two--pole 
ansatz with the two poles located at the same energies than the 
poles of the one--particle Green function. This treatment, 
beyond the spherical treatment of Nolting\cite{1}, can be mapped into the 
calculations of Kishore and Granato\cite{KG} with appropiate identification 
of our parameters in the paramagnetic phase. Their approach gives a Mott 
Metal--Insulator transition. All this means that the metal--insulator 
transition (MMIT) is embedded into the Hubbard model and it does not 
require of lifetime effects to accomplish this. However, once more, 
lifetime effects are a natural element of the many--body physics for 
intermediate and strongly correlated electron systems where the 
concept of quasi--particle does not apply any longer. We do not 
pursue anymore here the MMIT, but the interested reader is addressed 
to references\cite{9709080,Alexei,2,Varma}.

\indent By definition the one--particle Green function, 
$G(\vec{k},\omega)$, in terms of $\Sigma(\vec{k},\omega)$, is given as
\begin{equation}\label{1PGF}
G(\vec{k},\omega)  = \frac{1}{\omega - \varepsilon_{\vec{k}} 
- \Sigma(\vec{k},\omega)} ~~~ ,
\end{equation}
\noindent where $\varepsilon_{\vec{k}} = - 2t (\cos(k_x) 
+ \cos(k_y)  - \mu + \rho U$. Also, we will require the one--particle 
spectral density, $A(\vec{k},\omega)$, which is defined as
\begin{equation}\label{Akw}
A(\vec{k},\omega) = - \frac{1}{\pi} 
\lim_{\delta \rightarrow 0^+} Im 
G(\vec{k},\omega + i\delta) ~~~ 
\end{equation}

\indent Using Eqs. (\ref{single} - \ref{Akw}), we arrive to the 
following expression for the spectral density
\begin{equation}\label{11}
A(\vec{k},\omega) = \frac{-1}{\pi} 
\frac{\alpha(\vec{k}) \gamma(\vec{k})} 
{\left( (\omega - \varepsilon_{\vec{k}})(\omega - 
\Omega_{\vec{k}}) - \alpha(\vec{k}) \right)^2 + 
\gamma^2(\vec{k})(\omega - \varepsilon_{\vec{k}})^2 }
\end{equation}

\indent Using the first three sum rules of  Nolting\cite{1} 
for the the spectral function of Eq. (\ref{Akw}) we obtain 
the following equations
\begin{eqnarray}\label{3mom's}
\int_{-\infty}^{+\infty} A(\vec{k},\omega)d\omega &\equiv& 
M_0(\vec{k}) = 1 ~~~ \nonumber \\
\int_{-\infty}^{+\infty} \omega A(\vec{k},\omega)d\omega  
&\equiv& M_1(\vec{k}) = \varepsilon_{\vec{k}} ~~~
\nonumber \\
\int_{-\infty}^{+\infty} \omega^2 A(\vec{k},\omega)d\omega  
&\equiv& M_2(\vec{k}) = (\varepsilon_{\vec{k}}-\rho U)^2 + 
2\rho U (\varepsilon_{\vec{k}}-\rho U) + 
\rho U^2  
~~~ , 
\end{eqnarray}
\noindent where the $M_i(\vec{k})$'s, 
$i=0,1,2$, are the first three moments\cite{1}. For 
example, the first moment ($i = 0$) is the area below the curve of 
$A(\vec{k},\omega)$ vs $\omega$, the second 
moment ($i = 1$) is the center of gravity of the spectral 
function and the second order moment (or third moment, $i=2$) is 
related to the width of the spectral function, $A(\vec{k},\omega)$. 
So, damping effects are controlled by the second 
order sum rule. We do not use the fourth moment  
or sum rule because we have three $\vec{k}$--dependent 
unknown parameters (our way of working is different to one 
of Nolting since in the latter we have to use four moments. 
The difference lies in the fact that he starts with the one--particle 
spectral density). We could guess that in order to extend the 
canonical formalism of Nolting to include lifetime effects, 
starting from his two pole ansatz, we should have to postulate 
the following structure for $G(\vec{k},\omega)$
\begin{equation}\label{guessforGNolting}
G(\vec{k},\omega) = \frac{\alpha_1(\vec{k})}{\omega-\omega_1(\vec{k})
+i\gamma(\vec{k})} + \frac{\alpha_2(\vec{k})}{\omega-\omega_2(\vec{k})
+i\gamma(\vec{k})} ~~~,
\end{equation}
\noindent from where we see that we would need five moments or 
sum rules because we have five parameters to determine, i.e., 
$\alpha_i(\vec{k})$, $\omega_i(\vec{k})$, 
$\gamma(\vec{k})$, with $i=1,2$. With the proposal (Eq. (\ref{single})) 
we have only three parameters to calculate.
 
\indent We assume that at $\rho = 1/2$ the chemical 
potential, $\mu = U/2$. The density of states which results of 
the two pole Ansatz for the one--particle 
Green function, in the spherical 
approximation of Nolting\cite{1}, always has a gap. This 
solution (always a gap) is known in the literature as 
the Hubbard--I solution\cite{IS} which has been critized since 
many years ago by Laura Roth\cite{LR}, among others. We 
call the attention to Ref.\cite{oles} where the authors point out to the 
fact that the $\vec{k}$-dependence has to be included in 
$B(\vec{k})$.  A recent calculation 
by Kirchhofer\cite{diplome} is performed at the mean field level 
for the $\vec{k}$--dependence of the band narrowing 
factor, when the two Hubbard bands are separated. In a more 
elaborated calculation based on the Mori's formalism\cite{Mori} for the 
one--particle Green function, Kirchhoffer et al\cite{diplome,ours1} 
obtain three peaks in the spectral density, $A(\vec{k},\omega)$, which 
respects particle--hole symmetry. In the end, they get 
a Mott metal insulator transition, for $U/t = 5$.   
Here we are including  lifetime effects as a crucial ingredient in the 
formulation beyond a mean--field treatment. 
Kirchhofer\cite{diplome,ours1} also considers 
the presence of antiferromagnetism fluctuations in an empirical way. 
We could extend Kirchhoffer et al's calculations using the numerical 
values of the dynamical spin susceptibility, $\chi(\vec{q},\omega)$, 
in the spin--fermin model of superconductivity of Pines, Chubukov and 
others\cite{PinesChubukov}. The dynamical spin susceptibility has 
been obtained from nuclear magnetic resonance experiments in the 
High $T_c$ cuprates. In Section \ref{III}, 
we present our numerical results and their interpretation.

\section{Numerical results and their interpretation}\label{III}

\indent In Figs. 1a, 1b and 1c we present the {\it spectral density}, 
$A(\vec{k},\omega)$, vs $\omega$ along the diagonal of the Brillouin 
zone ($\vec{k} = 2\pi (n,n)/32$) for $U/W = 1/2,2/3$ and $1$, respectively. We 
are working with a finite system of periodicity of $32\times32$. For 
$U/W = 1/2$ we have a double--peaked structure, with visible 
lifetime effects (the Dirac delta functions of Nolting now have 
width). This is a feature of correlated electron systems as 
it has been discussed in the work of Schneider et al\cite{ours} for 
the case of $U < 0$. The physics is different but the peak structure 
is similar. For $U/W = 2/3$ we still observe the double peak structure 
but lifetime effects are stronger. Finally, for $U/W = 1$, the double 
peak structure is practically washed out. As we see, lifetime effects 
are very much pronounced for the larger values of $U/W$ presented, 
i.e., for $U/W = 1.0$. The two peaks of $A(\vec{k},\omega)$ vs 
$\omega$ are separated approximately by a distance of 
$\sqrt{(\varepsilon_{\vec{k}}-\Omega_{\vec{k}}-\gamma(\vec{k}))^2 
+ 4\gamma(\vec{k})\varepsilon_{\vec{k}}}$.

\indent In Figs. 2a, 2b and 2c we show the imaginary part of the 
self--energy, $-Im[\Sigma(\vec{k},\omega)]$, vs $\omega$  along the 
diagonal of the Brillouin zone for the same values of $U/W$ of 
Fig. 1. Again we observe that for increasing values of $U/W$, damping 
effects are stronger in the self--energy, as is the case in the 
one--particle spectral function (see Figs. 1a, 1b and 1c). In addition, 
we do not observe any Fermi liquid dependence (in frequency) of 
the imaginary part of the self--energy around $\omega = 0$. This 
is due to our choice of our ansatz (Eq. (\ref{single})). We could 
include Fermi or Marginal Fermi liquid behavior close to the 
chemical potential as it is suggested in Section \ref{IV}. However, 
within the present work, we could say that our approximation is 
valid for frequencies not too close to the chemical potential. Certaintly, 
for small values of $U/W$, we should have some Fermi liquid behavior 
(at least in 2--d), like an imaginary self--energy 
going to zero as positive power of $\omega$ at the chemical 
potential\cite{Metzner}.

\indent In Fig. 3a, 3b and 3c we report results for the 
real part of the self--energy, $Re[\Sigma(\vec{k},\omega)]$, vs 
$\omega$ along the diagonal of the Brillouin zone for the same 
values of interaction of Fig. 1. For $U/W = 1/2,2/3$ we see a more o less 
regular pattern. However, for $U/W = 1$, lifetime effects lead to big 
effects in $Re[\Sigma(\vec{k},\omega)]$.  For example, the curves are 
no longer regularly displaced with respect to one another. Also, the 
frequency range increases appreciably and the heights of the 
curves decrease. Let us comment that the numerical form of the real part 
of the self--energy clearly shows that the Kramers--Kronig 
relations for the self--energy are satisfied, in order 
to have the roots of the one--particle Green function on the 
same side of the complex plane.  Analitically, it 
can be proved too. The only requirement coming out of these  
calculations, with our Ansatz, is that $\alpha(\vec{k}) 
\geq 0$\cite{Matho} (see Appendix A). 
As $\alpha(\vec{k}) \times \gamma(\vec{k}) \leq 0$, 
then $\gamma(\vec{k}) \leq 0$. This considerations we have checked in 
further numerical calculations with lattice sizes of $64 \times 64$. 
Due to these new findings, we correct the results of Ref.\cite{9709080} 
since the solutions we found there must satisfy the conditions 
established here, i.e., $\alpha(\vec{k}) \geq 0$ and $\gamma(\vec{k}) 
\leq 0$.

\indent Figs. 4a, 4b and 4c show the imaginary part of the 
double-occupied Green function, $-Im[G_2(\vec{k},\omega)]$, 
vs $\omega$ along the diagonal of the Brillouin zone for the same values 
of interaction as before. Let us recall that $G_2(\vec{k},\omega)$ is given 
by 
\begin{equation}\label{twoG}
G_2(\vec{k},\omega)~ \equiv ~
\left<T_{\tau}\left[ c_{\vec{i},\sigma}(\tau)n_{\vec{i},\bar
{\sigma}}(\tau); c^{\dagger}_{\vec{j},\sigma}(0)\right] 
\right>_{(\vec{k},\omega)} ~~~ ,
\end{equation}
\noindent where $\bar{\sigma} = - \sigma$. 
In Eq. (\ref{twoG}) $(\vec{k},\omega)$ 
means the Fourier transform of the spatial--temporal correlation 
function and $T_\tau$ the usual time ordering of 
the operators. Using the equation of motion 
technique for the one--particle Green function, $G(\vec{k},\omega)$, we 
obtain that $G(\vec{k},\omega)$ and $G_2(\vec{k},\omega)$ are related as 
follows\cite{JJEAMF}
\begin{equation}\label{twoGE}
(\omega - \varepsilon_{\vec{k}})G(\vec{k},\omega) = 1 + U 
G_2(\vec{k},\omega)~~~ .
\end{equation}
\indent We observe that there is a big peak in the interval 
$\omega \in [-2,-1]$ 
which is most likely due to the peak in $A(\vec{k},\omega)$. However, the 
right peak at $(\vec{k},\omega/W) = (\pi,\pi,\approx 1/3)$ increases 
with interaction. At the same time, we see that the left frequency peaks 
($\omega < 0$) start to line up for small momenta but they almost vanish 
for $\vec{k} = \pi(3/4,3/4)$ and $\pi(1,1)$.
\begin{equation}\label{defchi2} 
\chi^{(2)}(\vec{k},\omega) \equiv 
-\frac{1}{\pi}\lim_{\delta \rightarrow 0^+}
Im[G_2(\vec{k},\omega+i\delta)] ~~~,
\end{equation}
\noindent is the spectral 
density for the double--occupied Green function. We see from 
Figs. 4 that there is 
negative contribution to this spectral density, which is due to the 
presence of the factor $\omega - \varepsilon_{\vec{k}}$ in front of the 
one--particle spectral density, $A(\vec{k},\omega)$. In addition, the  
factor $\omega - \varepsilon_{\vec{k}}$ is controlling the height 
of the peaks in $\chi^{(2)}(\vec{k},\omega)$. For example, when 
 $\chi^{(2)}(\vec{k},\omega) = 0$ is because this factor is zero. 
$\chi^{(2)}(\vec{k},\omega)$ is given by
\begin{equation}\label{chi2Akw}
U\chi^{(2)}(\vec{k},\omega) = (\omega-\varepsilon_{\vec{k}})\times 
A(\vec{k},\omega)~~~,
\end{equation}
\noindent which is identically zero for the non--interacting electron 
gas, since $A(\vec{k},\omega)$ is a Dirac delta function at the same 
argument of the quantity in front of it. So, any deviation from 
zero is a signature of an interacting system. 
Contrary to $A(\vec{k},\omega)$, 
which is always positive, $\chi^{(2)}(\vec{k},\omega)$ can be 
negative. The only requirement is that\cite{JJEAMF}
\begin{equation}\label{condition}
\int_{-\infty}^{+\infty}\chi^{(2)}(\vec{k},\omega) 
d\omega ~=~\rho ~~~ ,
\end{equation}
\noindent where $\rho$ is the electron density/spin. This can 
be easily checked calculating the first moment or moment of zeroth 
order for the double--occupied Green function. The relation 
between the self--energy and the double--occupied Green 
function is the following
\begin{equation}\label{Gamma_2Sigma}
UG_2(\vec{k},\omega) = \frac{\Sigma(\vec{k},\omega)}
{\left(\omega-\varepsilon_{\vec{k}}-\Sigma(\vec{k},\omega)\right)}~~~.
\end{equation}
\indent Eq. (\ref{Gamma_2Sigma}) is an exact relationship and 
it can be used to keep control of the approximations made in 
the self--energy and the double--occupied Green functions, as it 
has been discussed in Ref.\cite{paperanda}. Needless to say that 
to approximate $\Sigma(\vec{k},\omega)$ is equivalent to 
have an approximation for $G_2(\vec{k},\omega)$ and 
viceversa. In consequence, simple approximations for 
$G_2(\vec{k},\omega)$ are not always equivalent 
to simple approximations for 
$G(\vec{k},\omega)$ (or $\Sigma(\vec{k},\omega)$) or viceversa. 
For example, a single pole ansatz (without lifetime effects) in 
$G_2(\vec{k},\omega)$ leads to the Hubbard--I solution as it has been 
discussed in Ref.\cite{paperanda}. To go beyond the Hubbard--I 
solution for $G_2(\vec{k},\omega)$ we have to use Eqs. 
(\ref{1PGF},\ref{Gamma_2Sigma}).

\section{Conclusions and Future Trends}\label{IV}

\indent We have investigated the dynamical quantities, 
$A(\vec{k},\omega)$, $Re[\Sigma(\vec{k},\omega)]$, 
$-Im[\Sigma(\vec{k},\omega)]$ and $-Im[G_2(\vec{k},\omega)]$, 
vs $\omega$ along the diagonal of the Brillouin zone, for three 
values of the interaction, namely, $U/W = 1/2,2/3,1$. In all these 
quantities we observe that the role of correlations and lifetime 
effects is fundamental. For example, for values of $U/W \approx 
1$ the one--particle spectral density becomes almost one--peak, 
while $-Im[\Sigma(\vec{k},\omega)]$ becomes a wider inverted 
Lorentzian. $Re[\Sigma(\vec{k},\omega)]$, for $U/W = 1$ has lost 
all sign of regularity. $A(\vec{k},\omega)$ becomes  
featureless for large values of 
$U/W$. Our treatment of $G(\vec{k},\omega)$ and 
$G_2(\vec{k},\omega)$ is not perturbative since we impose 
sum rules to $A(\vec{k},\omega)$ to find $\Sigma(\vec{k},\omega)$ 
and $G_2(\vec{k},\omega)$ is found from the equation of motion 
technique (Eq. (\ref{twoGE})). 

\indent The choice of 
self--energy (Eq. (\ref{single}) is an attempt to shed some light on 
Nolting approach to which Eq. (\ref{single}) reduces when 
$\gamma(\vec{k}) = 0$\cite{RNMM}. Nolting's study (when looked 
upon with our optics, i.e., $\gamma(\vec{k}) = 0$, in Ref.\cite{RNMM}) is 
also a non--Fermi liquid. 
Our Ansatz for $\Sigma(\vec{k},\omega)$ 
is rather phenomenological, since we have not invoked any microscopic 
mechanism to postulate it (Eq. (\ref{single})). However, we have been guided 
by the single pole structure of Nolting without lifetime effects. This 
structure has been fleshed out in a recent paper\cite{RNMM}. 
Also, we have relied on the calculations of Kishore and 
Granato\cite{KG} which represent a non--Fermi 
liquid approach for the self--energy. Those interested in 
a nice interpretation of non--Fermi liquid behavior of the 
experimental data of  High--Temperature 
Cuprates, please see Ref.\cite{TT}. 
Work is in progress\cite{toappear} to include Fermi liquid 
features close to the chemical potential. According to our 
belief, this type of considerations are much harder to be tackled 
with the procedure of Nolting, i.e., two poles in 
the one--particle spectral function, $A(\vec {k},\omega)$. 
For example, the self--energy proposals of Norman et al\cite{DingRanderia},  
for the overdoped and underdoped regimes of the cuprate superconductors, 
can be numerically solved for the attractive Hubbard model\cite{Ventura}, 
for $d$--wave superconductivity, where off--diagonal Green 
functiond are called for.

\indent  A single pole structure in $\Sigma(\vec{k},\omega)$ 
goes beyond the Hubbard--I approximation, 
since the Hubbard--I approximation is also equivalent to choose a 
single pole in $G_2(\vec{k},\omega)$ (without 
lifetime effects). This can easily be checked due to the exact 
relationship given in Eq. (\ref{Gamma_2Sigma}). However, by 
comparing the results for the moments without lifetime effects we 
find that $\alpha(\vec{k}) \approx \rho (1-\rho)U^2$ which proves 
that our choice for $\Sigma(\vec{k},\omega)$ is, at least, a 
second order expansion in $U$. This is in agreement with the theoretical 
findings of Appendix A, since $\alpha(\vec{k}) \geq 0$. Thus, 
$\alpha(\vec{k})$ is almost $\vec{k}$--independent. Similarly, 
we find that $\gamma(\vec{k}) \leq 0$ is independent of $\vec{k}$, but 
strongly dependent on $U$. In consequence, our numerical study 
proves that our Ansatz is the easiest way to include lifetime 
effects and to consider Fermi and/or Marginal Fermi liquid 
behavior in the original proposal of the moment approach of 
Nolting.

    We thank  CONICIT--Venezuela (project F-139), 
the Brazilian Agency CNPq, FAPERGS and the Swiss National 
Science Foundation for finantial support.  
Interesting discussions with Prof. H. Beck, Prof. M.S. Figueira, Prof. E. 
Anda and Dr. M.H. Pedersen are fully acknowledged. In particular, 
Prof. Beck brought to our attention Ref.\cite{diplome}. 
Thanks to Mar\'{\i}a Dolores Garc\'{\i}a Gonz\'alez for 
a reading of the manuscript. 
%
%
%
%
\section{Appendix A: Poles of the One--Particle Green Function}\label{AA}
\indent With the self--energy ansatz given by Eq. (\ref{single}), the 
one--particle Green function becomes
\begin{equation}\label{A1}
G(\vec{k},\omega) = \frac{\omega-\Omega(\vec{k})-i\gamma(\vec{k})}
{(\omega-\varepsilon_{\vec{k}})(\omega-\Omega(\vec{k}))-\alpha(\vec{k})-
i\gamma(\vec{k})(\omega-\varepsilon_{\vec{k}})}~~~
\end{equation}
\indent From Eq. (\ref{A1}) the poles of the one--particle Green 
function are given by the roots of the following equation:
\begin{equation}\label{A2}
z^2 - (\varepsilon_{\vec{k}}+\Omega(\vec{k})+i\gamma(\vec{k}))z 
-\alpha(\vec{k}) + (\Omega(\vec{k})+i\gamma(\vec{k}))\varepsilon_{\vec{k}} = 0 
~~~.
\end{equation}
\indent Solving Eq. (\ref{A2}) we get that the two roots are
\begin{equation}\label{A3}
z_{\pm} = \frac{\varepsilon_{\vec{k}}+\Omega(\vec{k})+i\gamma(\vec{k}) 
\pm \sqrt{(\varepsilon_{\vec{k}}-\Omega(\vec{k})-i\gamma(\vec{k}))^2+
4\alpha(\vec{k})}}{2}
\end{equation}
\indent We have to find the real and imaginary parts of the two roots. 
For this we follow the standard procedure making
\begin{equation}\label{A4}
\sqrt{x+iy} = x_1 + iy_1~~~,
\end{equation}
\noindent from where we get that
\begin{equation}\label{A5}
x_1 = \left(\frac{\sqrt{x^2+y^2}+x}{2}\right)^2~~~; ~~~
y_1 = \left(\frac{\sqrt{x^2+y^2}-x}{2}\right)^2~~~.
\end{equation}
\indent Comparing Eqs. (\ref{A3}) and (\ref{A4}) we conclude
\begin{equation}\label{A6}
x \equiv (\varepsilon_{\vec{k}}-\Omega(\vec{k}))^2 + 4\alpha(\vec{k}) 
-\gamma^2(\vec{k})~~~;~~~y \equiv 2\gamma(\vec{k})(\Omega(\vec{k})-
\varepsilon_{\vec{k}}~~~.
\end{equation}
\indent In consequence, $z_{\pm}$ are given by
\begin{equation}\label{A7}
z_{\pm} = \frac{\Omega(\vec{k})+\varepsilon_{\vec{k}} \pm x_1 + 
i(\gamma(\vec{k}) \pm y_1)}{2}~~~.
\end{equation}
\indent If we require that our roots be on the upper half--complex 
plane, we must impose that $\gamma(\vec{k}) \pm y_1 \geq 0$. Carrying 
out the calculations we arrive to the result that $\alpha(\vec{k}) \geq 0$, 
which proves the statement advanced in Section \ref{III}.

\newpage
\vspace{1.6cm}
\begin{center}
{\huge FIGURES}
\end{center}

\vspace{0.8cm}
\noindent
Figs 1a, 1b and 1c. $A(n,n,\omega)$ vs $\omega$ along the 
diagonal of the Brillouin zone for three different 
values of interaction, namely, $U/W = 1/2,2/3$ and 
$1$. Our system has a periodicity of $32\times 32$. 
We are at half--filling, $\rho = 1/2$. As we work in 
two dimensions, the bandwidth is $W = 8t$. The wave 
vector along the diagonal is defined as $\vec{k} 
\equiv \frac{2\pi}{32}(n,n)$.

\vspace{0.6cm}
\noindent 
Figs. 2a, 2b and 2c. $-Im\Sigma(n,n,\omega)]~vs~\omega$ along 
the diagonal of the Brillouin zone. Same parameters 
of Fig. 1. 

\vspace{0.6cm}
\noindent  
Figs 3a, 3b and 3c. $Re[\Sigma(n,n,\omega)]~vs~n$ along the 
diagonal of the Brillouin zone.  Same parameters as previously.

\vspace{0.6cm}
\noindent Figs. 4a, 4b and 4c. $-Im[G_2(n,n,\omega)]~vs~\omega$ 
along the diagonal of the Brillouin zone. Same parameters as 
before. 

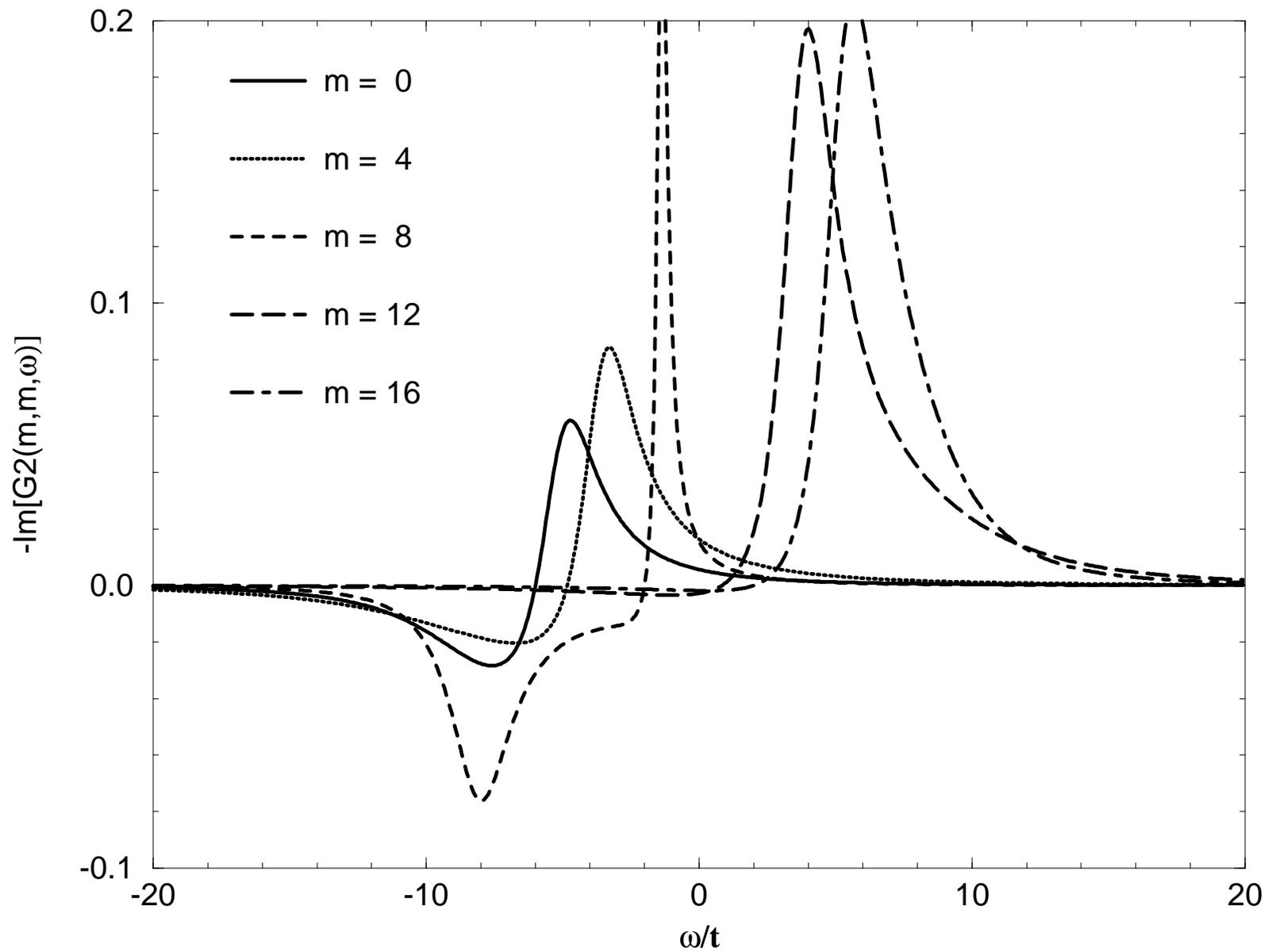
\end{document}